\documentclass[aps,twocolumn,prl,floatfix,groupedaddress,nofootinbib,notitlepage,showpacs,superscriptaddress]{revtex4-2}
\usepackage{times,bm,bbm,bbold,amssymb,amsmath,amsfonts,graphicx,graphics,hyperref,color,xcolor}
\hypersetup{colorlinks,linkcolor={blue},citecolor={blue},urlcolor= {blue}}

\DeclareFontFamily{OT1}{pzc}{}
\DeclareFontShape{OT1}{pzc}{m}{it}
{<-> s * [1.25] pzcmi7t}{}
\DeclareMathAlphabet{\mathpzc}{OT1}{pzc}
{m}{it}
\usepackage[T1]{fontenc} 
\usepackage{newtxtext,newtxmath} 

\begin{document}
	
	\title{Adaptive Quantum Heat Engines}
	\author{M. Khanahmadi}
	\affiliation{Department of Physics, Institute for Advanced Studies in Basic Sciences, Zanjan 45137, Iran}
	
	\author{A. T. Rezakhani}
	\email{rezakhani@sharif.edu}
	\affiliation{Department of Physics, Sharif University of Technology, Tehran 14588, Iran}
	\affiliation{School of Physics, Institute for Research in Fundamental Sciences (IPM), Tehran 19538, Iran}
	
	\begin{abstract}
		For heat engines working between two heat baths, functionality is often conditioned on a set of fixed constraints such as given internal structure of the engine and given temperatures for the baths. It is, however, important to devise heat engines which can function adaptively, in particular when the engine is a quantum system and the baths are subject to fluctuations. Here we study a model for an adaptive quantum heat engine whose heat baths can have variable temperatures. We obtain conditions under which such an engine can still operate. Moreover, we propose an enhancement of the heat engine by coupling it with an appropriate controller which changes the internal structure of the engine. Interestingly, we prove that this enhanced engine can always operate and we also obtain conditions for maximum power extraction from this engine for all temperatures of the heat baths. 
	\end{abstract}
	
	
	\date{\today}
	
	\maketitle
	
	\textit{Introduction.}---Since the inception of thermodynamics, heat engines (machines operating between two reservoirs with different temperatures and converting heat to work) have been a cornerstone concept \cite{Callen}. Recently with the renewed interest in thermodynamics in quantum regimes \cite{q-thermo-1, q-thermo-2, q-thermo-3, q-thermo-4}, investigation of quantum heat engines has also been reinvigorated \cite{Alicki, Uzdin, Geusrc, Kosloff, Scovil, Scully1, Kosloff-review, Esposito, Arnab, Konstantin, Kosloff1, Tannor, Linden, Linden-PRL, Artur, Wolfgang, allahverdyan2, Ramezani}, e.g., operation of \textit{autonomous} heat engines has been extensively analyzed \cite{Linden-PRL, Artur, Wolfgang, allahverdyan2, Frenzel, Friedemann,Roulet,Verteletsky}.
	
	A heat engine optimal or close-to-optimal for some given conditions may not necessarily function optimally when its ambient conditions change. For example, in realistic heat engines temperatures of heat baths may have inevitable fluctuations. Thus, it may be important so see whether heat engines can adapt themselves with such variations and continue to function under new conditions \cite{allahverdyan,allahverdyan1}. This issue will be more important for small-scale heat engines working in quantum regimes. Here we propose an \textit{adaptive} quantum heat engine which can extract energy with maximum power on the work source, at different temperatures of its baths.
	
	\begin{figure*}[tp]
		\includegraphics[scale=.32]{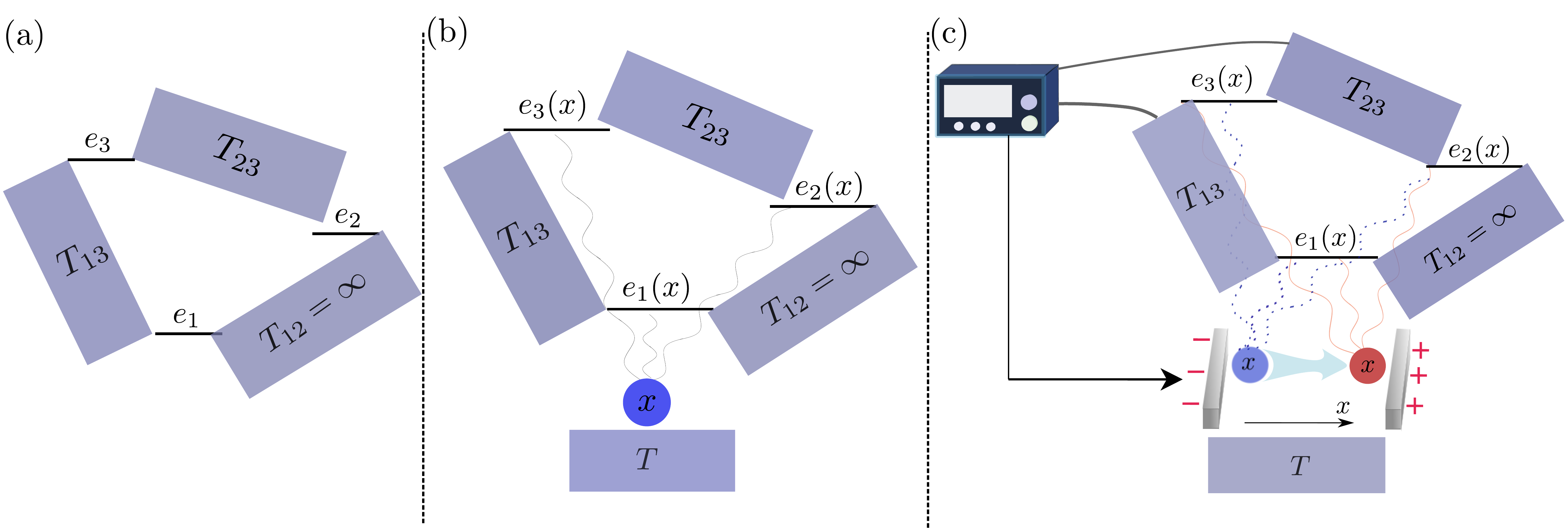}
		\caption{Schematic of a three-level quantum heat engine system (a) coupled to three thermal baths with different temperatures and a controller (b). (c) Schematic of our adaptive framework. A programmed \textit{learner} with two thermometers measures temperatures of the hot and cold baths and computes the stationary position $x^{*}$ of the controller, the conditional work $J_{12|x}$ of the heat engine, and the position $\tilde{x} = \mathrm{argmax}\,|J_{12|x}|$. Next, the learner compares $x^{*}$ and $\tilde{x}$; if $x^{*}\neq \tilde{x}$, the learner computes the proper electric field $E'$ from Eq. (\ref{ee1}) and applies it to the controller to move it to the position where the heat engine can operate with its maximum power.}
		\label{f1}
	\end{figure*}
	
	\textit{Model}.---We consider a minimal model for a quantum heat engine, a three-level quantum system which interacts with two heat baths \cite{Scovil, Scully1, Kosloff-review, Esposito, Arnab, Konstantin, Kosloff1, Tannor, Linden, Linden-PRL}. The Hamiltonian of the system is $H_{0}=\sum_{i=1}^{3} e_{i} |i\rangle \langle i |$, where $e_{i}$s are the energies and $\{|i\rangle\}$ are energy eigenstates. Transitions between the energy levels $i$ and $j$ are induced by exchanging energy with a thermal bath of temperature $T_{ij}$, mediated by an appropriate frequency filter passing only $\omega_{ij}=e_{j}-e_{i}$ \cite{Scovil} (assuming $\hbar \equiv1$ hereafter)---Fig. \ref{f1}. The thermal baths with finite temperatures $T_{23}$ and $T_{13}$ are considered as the \textit{cold} and \textit{hot} reservoirs (depending which temperature is higher or lower). However, the temperature of the thermal bath $T_{12}$ is assumed to be infinity, which renders it effectively as a \textit{work source} \cite{Esposito}. 
	
	We assume that the dynamics is Markovian and described by the local Lindblad master equation (disregarding the Lamb shift) \cite{Breuer,DeChiara}
	\begin{align}
	\dot{\varrho}_{S} = -i[H_{0},\varrho_{S}] + \textstyle{\sum_{j>i=1}^{3}} \mathpzc{L}_{ij} [\varrho_{S}] ,
	\label{t1}
	\end{align}
	where dot denotes time derivative, $\varrho_{S}$ is the state of the system, and $\mathpzc{L}_{ij}$ is the local Lindbladian term caused by interaction with the thermal bath $ij$ (typically assumed a bath of quantized radiation fields) given by \cite{Breuer}
	\begin{align}
	\label{e2}
	\mathpzc{L}_{ij} [\circ]=&\, \gamma_{i\leftarrow j}[S_{ij} \circ S^{\dagger}_{ij}-\frac{1}{2}\lbrace S^{\dagger}_{ij}S_{ij}, \circ \rbrace ]
	\\ \nonumber &+ \gamma_{j\leftarrow i}[S^{\dagger}_{ij} \circ S_{ij}-\frac{1}{2}\lbrace S_{ij}S^{\dagger}_{ij},\circ \rbrace ],
	\end{align}
	where $S_{ij}=|i\rangle\langle j|$ are jump operators with rates
	\begin{align}
	\label{t2}
	\gamma_{i\leftarrow j} =\begin{cases}
	\gamma_{0}|\omega_{ij}|^{3} e^{\beta_{ij}|\omega_{ij}| } (e^{\beta_{ij}|\omega_{ij}|}-1)^{-1},& j>i\\
	\gamma_{0} |\omega_{ij}|^{3} (e^{\beta_{ij}|\omega_{ij}| }-1)^{-1},& i>j
	\end{cases}
	\end{align}
	and $\beta_{ij}=1/(k_{B}T_{ij})$ and $\gamma_{0}>0$. The average energy $U(t)=\mathrm{Tr}[H_{0}\varrho_{S}(t)]$ changes in time as
	\begin{align}
	\label{av}
	\dot{U}(t)= J_{12}(t)+J_{13}(t)+J_{23}(t),
	\end{align}
	where the average rate of energy exchange $J_{ij}(t)$ ($ij\in\{12,23,13\}$) between the system and each thermal bath is $J_{ij}(t) = [\gamma_{i\leftarrow j}p_{j}(t)-\gamma_{j\leftarrow i}p_{i}(t)](e_{i}-e_{j})$, with $p_{i}(t)=\langle i|\varrho_{S}(t)|i\rangle$ being the population of the level $|i\rangle$. When $J_{ij} >0$ ($J_{ij}<0$) the system absorbs (releases) energy from (into) bath $T_{ij}$. In the \textit{stationary state} $\varrho^{(\mathrm{ss})}$, given by the populations
	\begin{align}
	\label{sp}
	p^{(\mathrm{ss})}_{i}=(1/Z)[\gamma_{i\leftarrow j}\gamma_{i\leftarrow k}+\gamma_{i\leftarrow j}\gamma_{j\leftarrow k}+\gamma_{i\leftarrow k}\gamma_{k\leftarrow j}],
	\end{align}
	with $i\in\{1,2,3\}$, $i\neq j\neq k$, and $Z$ ensuring normalization, we have $\sum_{j>i} J^{(\mathrm{ss})}_{ij}=0 $. The power of the engine is given by $P = \dot{W}$, where $W$ is the work done by the engine, which is equal to the heat coming from the bath $T_{12}$. In the stationary state (dropping hereafter superscript ``($\mathrm{ss}$)'' for brevity) we have 
	\begin{equation}
	\label{tt1}
	P = J_{12} = \gamma_{12}(p_{1}-p_{2})(e_{2}-e_{1}),
	\end{equation}
	where we have used $\gamma_{1\leftarrow 2} \approx \gamma_{2\leftarrow 1} =:\gamma_{12}$ (due to $T_{12}\rightarrow \infty$). By introducing $\hat{e}_{i}\equiv e_{i}-e_{1}$ ($i\neq 1$) we obtain
	\begin{align}
	\label{t16}
	J_{12} =\frac{\hat{e}_{2}}{Z}\gamma_{12}\gamma_{1\leftarrow 3}\gamma_{3\leftarrow2}[1-e^{(\beta_{32}-\beta_{31})\hat{e}_{3}-\beta_{32}\hat{e}_{2}}]\equiv \hat{e}_{2}\mathbbmss{j}.
	\end{align}
	The heat engine operates if $ J_{12}<0 $, hence as long as $\theta \equiv T_{32}/T_{13}$ satisfies
	\begin{align}
	\label{ad}
	-\hat{e}_{2}[(1-\theta)\hat{e}_{3}-\hat{e}_{2}]<0,
	\end{align}
	the engine can \textit{adapt} itself with the change in the temperatures of the heat baths. 
	
	If the temperatures vary such that the adaptability condition does not hold, one remedy can be to change the energy levels $e_{i}$ appropriately. This can be achieved by coupling a controller to the system, whose effect is to modify the system energy levels depending on its degree of freedom. We assume that the controller $C$ is a quantum Brownian particle of mass $m$ with $(x,p)$ position-momentum degrees of freedom and potential $(1/2) \kappa x^{2}$, coupled to another thermal bath of temperature $T$, which is described by the Caldeira-Leggett model \cite{Breuer,x}. The dynamics of the state of the controller $\varrho_{C}(t)$ is given by $\dot{\varrho}_{C}=-i[H_{C},\varrho_{C}]+\mathpzc{L}_{\mathrm{CL}}[\varrho_{C}]$, where $H_{C}= (1/2m) p^{2} + (1/2)\kappa x^{2}$ is the Hamiltonian of the controller and $\mathpzc{L}_{\mathrm{CL}}[\circ] = -i\xi[x,\lbrace p,\circ\rbrace]-2m\xi T[x,[x,\circ]]$, with $\xi$ being the friction coefficient and $\{A,B\}\equiv AB+BA$. The stationary state probability density of the controller is given by $p_{C}(x)=\sqrt{\kappa/(2\pi T) } e^{-\kappa x^{2}/(2T)}$. Let us introduce 
	\begin{equation}
	x^{*} \equiv \mathrm{argmax}\,p_{C}(x), 
	\end{equation}
	as the most probable position of the controller in the stationary state. We consider the system-controller interaction as \cite{Kovarskii}
	\begin{equation}
	\label{c11}
	H_{SC}= \textstyle{\sum_{i=1}^{3}} g_{i}|i\rangle \langle i|\otimes x,
	\end{equation}
	where the controller is coupled to the levels $|i\rangle$ with different couplings $g_{i}$. Hence the state of the system-controller in the weak-coupling, Markovian regime is given by 
	\begin{align}
	\label{e1}
	\dot{\varrho}_{SC} &= -i[H_{0}+H_{C}+H_{SC},\varrho_{SC}] + (\mathpzc{L}_{\mathrm{CL}} + \textstyle{\sum_{j>i}} \mathpzc{L}_{ij}) [\varrho_{SC}].
	\end{align}
	
	\begin{figure*}[tp]
		\includegraphics[scale=.4]{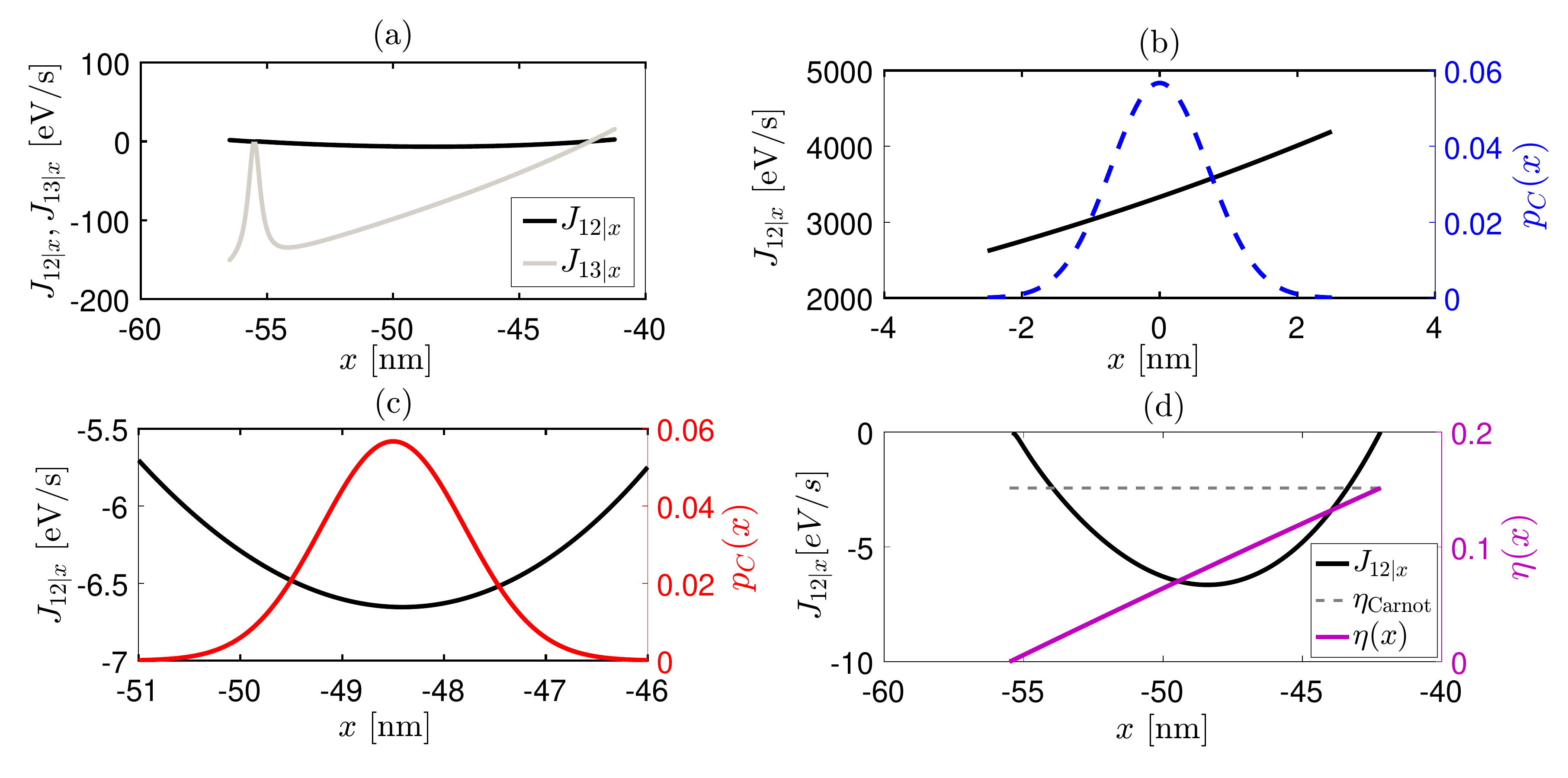}
		\caption{An example of an adaptive quantum heat engine: (a) Conditional power of the engine and heat current of the hot thermal bath vs. the position of the controller. The minimum power is at $\tilde{x}\approx -48.5\,[\mathrm{nm}]$. (b) Stationary probability of the controller with $ x^{*}\approx 0\,[\mathrm{nm}]$ is shown with blue chart. At this position the conditional work (black line) is positive so it is needed to apply some proper electric force $E$ to move the controller to the position at which the engine operates with maximum power; shown in panel (c). (d) Conditional efficiency \eqref{add}. The dashed line shows $\eta_{\mathrm{Carnot}}$. To generate these plots, the values of the model parameters are chosen as follows: $\kappa =10^{-12}\, [\mathrm{N}/\mathrm{nm}]$, $e_{1}=-5.2\,[\mathrm{eV}]$, $e_{2}=-3.4\,[\mathrm{eV}]$, $e_{3}=-1.2\, [\mathrm{eV}]$, $T_{13}=330 \, [\mathrm{K}]$, $T_{23}=T=280\, [\mathrm{K}]$, $(g_{1},g_{2},g_{3})=(1.77, 2.16, 1.87)\times10^{-3}\,[\mathrm{eV}/\mathrm{nm}]$, $|E|\approx 10^{-11}\, [\mathrm{N}]$, $\xi = 10^{-10}\, [1/\mathrm{nm}]$, and $m =10^{-22}[\mathrm{g}]$.}
		\label{f3}
	\end{figure*}
	The controller affects the energy levels of the system and hence its functioning as the heat engine. In order to find the stationary state of the controller and $x^{*}$, one should use Eq. (\ref{e1}). To obtain the engine's power associated with the position of the controller, we need to find the conditional state $\varrho_{S|C}=\mathrm{Tr}_{C}[\mathbbmss{I}_{S}\otimes |x\rangle_{C}\langle x|\, \varrho_{SC}]$, which obeys the master equation
	\begin{align}
	\label{a1}
	\dot{\varrho}_{S|C}= -i[H(x),\varrho_{S|C}]+ \textstyle{\sum_{j>i}} \mathpzc{L}_{ij|C} [ \varrho_{S|C}],
	\end{align}
	where \cite{Kosloff1}
	\begin{equation}
	H(x) = \textstyle{\sum_{i}}  e_{i}(x) |i\rangle \langle i|= \textstyle{\sum_{i}} (e_{i}+g_{i}x)|i\rangle \langle i|
	\end{equation}
	and $\mathpzc{L}_{ij|x} $ has the form as in Eq. (\ref{e2}) with $S_{ij|x}=S_{ij} $ and $\gamma_{i\leftarrow j} \to\gamma_{i\leftarrow j} (x)$ as in Eq. (\ref{t2}) with the difference that now $\omega_{ij}\to \omega_{ij}(x) \equiv e_{j}(x)- e_{i}(x)$. The stationary solution of Eq. (\ref{a1}) yields populations as in Eq. (\ref{sp}) where all $\gamma_{ab}\to \gamma_{ab}(x)$. Thus, from Eq. (\ref{tt1}) the conditional power is obtained as 
	\begin{align}
	\label{aa10}
	J_{12|x} =\hat{e}_{2}(x)\,\mathbbmss{j}(x),
	\end{align}
	and similalry $ J_{13|x}= -\hat{e}_{3}(x)\, \mathbbmss{j}(x) $ and $  J_{23|x}= [\hat{e}_{3}(x)-\hat{e}_{2}(x)] \, \mathbbmss{j}(x) $, where $\hat{e}_{i}(x) \equiv   e_{i}(x)-e_{1}(x)$. In this scenario, the adaptation condition $J_{12|x}<0$ yields
	\begin{align}
	\label{aa15}
	-\hat{e}_{2}(x)[(1-\theta)\hat{e}_{3}(x)-\hat{e}_{2}(x)]<0, \,\,\,\,\,\, x\approx x^{*}
	\end{align}
	which means that the most probable position of the controller should be where the heat engine can perform work on the work source. For later use, here we introduce a position $\tilde{x}$ where 
	\begin{equation}
	\tilde{x} \equiv \mathrm{argmax}\,|J_{12|x}|, 
	\end{equation} 
	at which the extracted power from the engine in optimal. It is ideal to have $\tilde{x} = x^{*}$. Satisfying Eq. (\ref{aa15}) reduces to making the quadratic form $y(x) \equiv ax^{2}+bx+c$ negative, where $a =(\theta-1)\hat{g}_{2}\hat{g}_{3}+\hat{g}_{2}^{2}$, $b = (\theta-1)(\hat{e}_{2}\hat{g}_{3}+\hat{g}_{2}\hat{e}_{3})+2\hat{g}_{2}\hat{e}_{2}$, $c = (\theta-1)\hat{e}_{3}\hat{e}_{2}+\hat{e}^{2}_{2}$, and $\hat{g}_{i} = g_{i}-g_{1}$. Note that the discriminant of $y(x)$ is always nonnegative, 
	\begin{align}
	\label{delta}
	[(1-\theta) (\hat{e}_{2}\hat{g}_{3}-\hat{g}_{2}\hat{e}_{3})]^{2}\geqslant 0.
	\end{align}
	Depending on $x^{*}$, the sign of $y(x^{*})$ can be positive (unacceptable) or negative (acceptable). However, it is interesting that according to Eq. (\ref{delta}), and provided that $ \hat{e}_{2}\hat{g}_{3}\neq \hat{e}_{3}\hat{g}_{2}$, at all temperature of the hot and cold baths, one can always find acceptable ranges for $x$. In other words, in contrast to the case of \textit{classical} heat engines \cite{allahverdyan}, by attaching a controller to a \textit{quantum} heat engine we can always make it adaptive.
	
	We can also calculate the efficiency of the heat engine as
	\begin{align}
	\label{add}
	\eta(x) = \frac{-J_{12|x}}{\mathrm{max}\{J_{23|x},J_{13|x}\}} \leqslant \eta_{\mathrm{Carnot}} = 1-\mathrm{min}\{\theta,1/\theta\}, 
	\end{align}
	depending on whether $T_{13}>T_{23}$ or $T_{23}>T_{13}$. The efficiency is upper bounded by the Carnot efficiency $\eta_{\mathrm{Carnot}}$, and when the heat engine reaches its maximum efficiency, according to Eq. \eqref{aa15} its power goes to zero---so that it respects the trade-off between power and efficiency \cite{Callen}. From Fig. \ref{f3} (a) it is seen that the power vanishes at two points $x_{0} = -55.5\,[\mathrm{nm}]$ and $x_{1}=-42.1\,[\mathrm{nm}]$, and that in the region $(x_{0},x_{1})$ the heat engine can perform work on its environment. Note that $\eta (x_{0}) = 0$ and $\eta (x_{1}) = 1-\theta = \eta_{\mathrm{Carnot}}$. This can be explained by noting that when the controller is at $x_{0}$, we have degeneracy and the energy levels $|1\rangle$ and $|2\rangle$ coincide [$\hat{e}_{2}(x_{0})= 0$], hence no work can be extracted from the heat engine, whereby the power and the efficiency both vanish. However, at $x_{1}$ the three levels are degenerate [$\hat{e}_{2}(x_{1})=\hat{e}_{3}(x_{1})= 0$], which implies that in the vicinity of that position the power and the heat absorbed from the hot bath approach $0$ such that $\eta (x_{1})\to\eta_{\mathrm{Carnot}}$. 
	
	If $ y(x^{*})>0$, the system cannot operate as a heat engine. Thus, we need to move the controller to the acceptable region of $x$ as given by Eq. (\ref{aa15}). This can be achieved by assuming that the controller is a charged particle with charge $q$ and applying an appropriate electric field $E'$ on it.  In this case, the position of the controller is obtained from $\dot{\varrho}_{C}=-i[H_{C}+H_{CF},\varrho_{C}]+\mathpzc{L}_{\mathrm{CL}}[\varrho_{C}]$, where $H_{CF}=Ex$, with $E\equiv qE'$. This yields that the stationary probability density of the position of the controller is modified to $p_{C}(x) = [e^{-E^{2}/(2\kappa T)}\sqrt{\kappa/(2\pi T)}] e^{-(Ex+\kappa x^{2}/2)/T}$. Under the potential $H_{CF}$ the dynamical equation (\ref{e1}) changes to
	\begin{align}
	\label{ee1}
	\dot{\varrho}_{SC}=& -i[H_{0}+H_{C}+H_{SC}+H_{CF},\varrho_{SC}]\nonumber\\&+
	\mathpzc{L}_{\mathrm{CL}} [\varrho_{SC}]+\textstyle{\sum_{j>i}^{3}} \mathpzc{L}_{ij}[ \varrho_{SC}], 
	\end{align}
	from which one can see that the same conditional master equation (\ref{a1}) and power (\ref{aa10}) are obtained; hence the profile of the conditional power does not change by the electric field. 
	
	By attaching a programmed \textit{learner} to the engine, we can simply make its adaptation autonomous and autorun (at the price of energy needed to keep the learner operating). Before applying the electric filed, the learner calculates the position $\tilde{x}$ and compares it with $x^{*}$; then it turns on a proper electric field on the charged controller to enforce $x^{*}=\tilde{x}$---Fig. \ref{f3}. This way the engine can operate adaptively and with maximum power at different temperatures of its baths---Fig. \ref{f1}. 
	
	\textit{Summary and conclusions.}---Here we have investigated an adaptive model for a quantum heat engine whose baths may have nonconstant temperatures. The pivotal ingredient of the model is coupling it with a charged quantum Brownian controller whose position can always be chosen such that the engine can perform work on its environment. We have shown that this adaptation can be performed in an autorun fashion with the assistance of a programmed learner which forces the controller to move to a position at which the engine can operate with maximum power. 
	
	\textit{Acknowledgments.}---Partially supported by Sharif University of Technology's Office of Vice President for Research and Technology, School of Nano Science, and School of Physics at the Institute for Research in Fundamental Sciences (IPM).
	
	
	

\begin{thebibliography}{100}
		
		\bibitem{Callen} H. B. Callen, \emph{Thermodynamics} (John Wiley, New York, 1985).
		
		\bibitem{q-thermo-1} J. Gemmer, M. Michel, and G. Mahler, \emph{Quantum Thermodynamics -- Emergence of Thermodynamic Behavior within Composite Quantum Systems} (Springer, Berlin, 2009).
		
		\bibitem{q-thermo-2} F. Binder, L. Correa, C. Gogolin, J. Anders, and G. Adesso (Eds.), \emph{Thermodynamics in the Quantum Regime -- Fundamental Aspects and New Directions} (Springer International, Cham, Switzerland, 2018).

		\bibitem{q-thermo-3} S. Alipour, F. Benatti, F. Bakhshinezhad, M. Afsary, S. Marcantoni, and A. T. Rezakhani, Correlations in quantum
thermodynamics: Heat, work, and entropy production, \href{https://doi.org/10.1038/srep35568}{Sci. Rep. \textbf{6}, 35568 (2016)}.
	
	          \bibitem{q-thermo-4} M. Perarnau-Llobet, H. Wilming, A. Riera, R. Gallego, and J. Eisert, Strong Coupling Corrections in Quantum Thermodynamics, \href{https://doi.org/10.1103/PhysRevLett.120.120602}{Phys. Rev. Lett. \textbf{120}, 120602 (2018)}. 
	
		\bibitem{Scovil} H. E. D. Scovil and E. O. Schultz-DuBois, Three-level masers as heat engines, \href{https://doi.org/10.1103/PhysRevLett.2.262}{Phys. Rev. Lett. \textbf{2}, 262 (1959)}.
		
		\bibitem{Geusrc}J. E. Geusic, E. O. Schulz-DuBios, and H. E. D. Scovil, Quantum Equivalent of the Carnot Cycle, \href{https://doi.org/10.1103/PhysRev.156.343}{Phys. Rev. \textbf{156}, 2 (1967).}
		
		\bibitem{Alicki}R. Alicki, The quantum open system as a model of the heat engine, \href{https://doi.org/10.1088/0305-4470/12/5/007}{J. Phys. A \textbf{12}, L103 (1979).}
		
		\bibitem{Kosloff} R. Kosloff, A quantum mechanical open system as a model of a heat engine, \href{https://doi.org/10.1063/1.446862}{J. Chem. Phys. \textbf{80}, 1625 (1984)}. 
		
		\bibitem{Scully1} M. O. Scully, K. R. Chapin, K. E. Dorfman, M. B. Kim, and A. Svidzinsky, Quantum heat engine power can be increased by noise-induced coherence, \href{https://doi.org/10.1073/pnas.1110234108}{Proc. Natl. Acad. Sci. U.S.A. \textbf{108}, 15097 (2011)}.

		\bibitem{Kosloff-review} R. Kosloff and A. Levy, Quantum heat engines and refrigerators: Continuous devices \href{https://doi.org/10.1146/annurev-physchem-040513-103724}{Annu. Rev. Phys. Chem. \textbf{65}, 365 (2014)}.
		
		\bibitem{Uzdin}R. Uzdin, A. Levy, and R. Kosloff, Equivalence of Quantum Heat Machines, and Quantum-Thermodynamic Signatures, \href{https://doi.org/10.1103/PhysRevX.5.031044}{Phys. Rev. X \textbf{5}, 031044 (2015).}
				
		\bibitem{Esposito} M. Esposito, K. Lindenberg, and C. Van den Broeck, Universality of Efficiency at Maximum Power, \href{https://doi.org/10.1103/PhysRevLett.102.130602} {Phys. Rev. Lett. \textbf{102}, 130602 (2009)}.
		
		\bibitem{Arnab} A. Ghosha, D. Gelbwaser-Klimovsky, W. Niedenzu, A. I. Lvovsky, I. Mazets, M. O. Scully, and G. Kurizki, Two-level masers as heat-to-work converters, \href{https://doi.org/10.1073/pnas.1805354115}{Proc. Natl. Acad. Sci. U.S.A. \textbf{115}, 9941 (2018).}
		
		\bibitem{Konstantin} K. E. Dorfmana, D. V. Voroninea, S. Mukamel, and M. O. Scully, Photosynthetic reaction center as a quantum heat engine, \href{https://doi.org/10.1073/pnas.1212666110}{Proc. Natl. Acad. Sci. U.S.A. \textbf{110}, 27468 (2013)}.
		
		\bibitem{Kosloff1} E. Geva and R. Kosloff, Three-level quantum amplifier as a heat engine: A study in finite-time thermodynamics, \href{https://doi.org/10.1103/PhysRevE.49.3903}{Phys. Rev. E \textbf{49}, 3903 (1994)}. 
		
		\bibitem{Tannor} E. Boukobza and D. J. Tannor, Three-Level Systems as Amplifiers and Attenuators: A Thermodynamic Analysis, \href{https://doi.org/10.1103/PhysRevLett.98.240601}{Phys. Rev. Lett. \textbf{98}, 240601 (2007)}.
	
		\bibitem{Linden} N. Linden, S. Popescu, and P. Skrzypczyk, The smallest possible heat engines, \href{http://arxiv.org/abs/1010.6029}{arXiv:1010.6029 (2010)}.
		
		\bibitem{Ramezani} M. Ramezani, S. Marcantoni, F. Benatti, R. Floreanini, F. Petiziol, A. T. Rezakhani, and M. Golshani, Impact of nonideal cycles on the efficiency of quantum heat engines, \href{https://doi.org/10.1140/epjd/e2019-90520-7}{Eur. Phys. J. D \textbf{73}, 144 (2019)}.
		
		\bibitem{Linden-PRL} N. Linden, S. Popescu, and P. Skrzypczyk, How Small Can Thermal Machines Be? The Smallest Possible Refrigerator, \href{https://doi.org/10.1103/PhysRevLett.105.130401}{Phys. Rev. Lett. \textbf{105}, 130401 (2010)}.
			
		\bibitem{Artur} A. S. L. Malabarba, A. J. Short, and P. Kammerlander, Clock-driven quantum thermal engines, \href{https://doi.org/10.1088/1367-2630/17/4/045027}{New J. Phys. \textbf{17}, 045027 (2015)}.
		
		\bibitem{Wolfgang} W. Niedenzu, M. Huber, and E. Boukobza, Concepts of work in autonomous quantum heat engines, \href{https://doi.org/10.22331/q-2019-10-14-195}{Quantum \textbf{3}, 195 (2019)}. 
		
		\bibitem{allahverdyan2} A. E. Allahverdyan, R. Balian, and Th. M. Nieuwenhuizen, Maximal work extraction from finite quantum systems, \href{https://doi.org/10.1209/epl/i2004-10101-2}{Europhys. Lett. \textbf{67}, 565 (2004)}.
				
		\bibitem{Frenzel} M. F. Frenzel, D. Jennings, and T. Rudolph, Quasi-autonomous quantum thermal machines and quantum to classical energy flow, \href{https://dx.doi.org/10.1088/1367-2630/18/2/023037}{New J. Phys. \textbf{18}, 023037 (2016)}.
		
		\bibitem{Friedemann} F. Tonner and G. Mahler, Autonomous quantum thermodynamic machines, \href{https://doi.org/10.1103/PhysRevE.72.066118}{Phys. Rev. E \textbf{72}, 066118 (2005)}.
		
		\bibitem{Roulet} A. Roulet, S. Nimmrichter, J. M. Arrazola, S. Seah, and V. Scarani, Autonomous rotor heat engine, \href{https://doi.org/10.1103/PhysRevE.95.062131}{Phys. Rev. E \textbf{95}, 062131 (2017)}.
		
		\bibitem{Verteletsky}K. Verteletsky and K. M{\o}lmer, Revealing the strokes of autonomous quantum heat engines with work and heat fluctuations, \href{https://doi.org/10.1103/PhysRevA.101.010101}{Phys. Rev. A \textbf{101}, 010101 (2020)}.
		
		\bibitem{allahverdyan} A. E. Allahverdyan, S. G. Babajanyan, N. H. Martirosyan, and A. V. Melkikh, Adaptive Heat Engine, \href{https://doi.org/10.1103/PhysRevLett.117.030601}{Phys. Rev. Lett. \textbf{117}, 030601 (2016)}.
		
		\bibitem{allahverdyan1} A. E. Allahverdyan and Q. A. Wang, Adaptive machine and its thermodynamic costs, \href{https://doi.org/10.1103/PhysRevE.87.032139}{Phys. Rev. E \textbf{87}, 032139 (2013)}.
		
		\bibitem{Breuer} H.-P. Breuer and F. Petruccione, \emph{The Theory of Open Quantum Systems} (Oxford University Press, New York, 2002).
		
		\bibitem{DeChiara} G. De Chiara, G. Landi, A. Hewgill, B. Reid, A. Ferraro, A. J. Roncaglia, and M. Antezza, Reconciliation of quantum local master equations with thermodynamics, \href{https://doi.org/10.1088/1367-2630/aaecee}{New J. Phys. \textbf{20}, 113024 (2018)}.
		
		\bibitem{x} A. O. Caldeira and A. J. Leggett, Path integral approach to quantum Brownian motion, \href{https://doi.org/10.1016/0378-4371(83)90013-4}{Physica A \textbf{121}, 587 (1983)}.
		
		\bibitem{Kovarskii}V. A. Kovarskii, Quantum processes in biological molecules. Enzyme catalysis, \href{https://doi.org/10.1070/PU1999v042n08ABEH000480}{Phys. Usp. \textbf{42}, 797 (1999)}.
		
%
		
	\end{thebibliography}
\end{document}